\documentclass{egpubl}

\usepackage[utf8]{inputenc}
\usepackage{amssymb}
\usepackage{multirow}
\JournalSubmission

\electronicVersion

\usepackage{graphicx}

\PrintedOrElectronic
\usepackage{egweblnk}
\usepackage{cite}
\usepackage{t1enc,dfadobe}

\title[A survey of ocean simulation and rendering techniques]{A survey of ocean simulation and rendering techniques in computer graphics}
\author[E. Darles \& B. Crespin \& D. Ghazanfarpour \& JC. Gonzato]
  {E. Darles$^1$, B. Crespin$^1$, D. Ghazanfarpour$^1$, JC. Gonzato$^2$\\
  $^1$XLIM - University of Limoges, France\\$^2$University of Bordeaux (LaBRI) \& INRIA, France}

\begin{document}

\maketitle

\begin{abstract}
This paper presents a survey of ocean simulation and rendering methods in computer graphics. To model and animate the ocean's surface, these methods mainly rely on two main approaches: on the one hand, those which approximate ocean dynamics with parametric, spectral or hybrid models and use empirical laws from oceanographic research. We will see that this type of methods essentially allows the simulation of ocean scenes in the deep water domain, without breaking waves. On the other hand, physically-based methods use Navier-Stokes Equations (NSE) to represent breaking waves and more generally ocean surface near the shore. We also describe ocean rendering methods in computer graphics, with a special interest in the simulation of phenomena such as foam and spray, and light's interaction with the ocean surface.
\end{abstract}



\section{Introduction}

The main goal of computer graphics is to reproduce in the most true-to-life possible way the perceived reality with all the complexity of natural phenomena surrounding us. The ocean's complexity is mainly due to a highly dynamic behavior. Ranging from a quiet sea to an agitated ocean, from small turbulent waves to enormous shorebreaks, the dynamic motion of the ocean is influenced by multiple phenomena occurring at small and large scales. For several centuries, scientists have tried to understand and explain these mechanisms. In oceanographic research, physicists define the behavior of the ocean surface depending on its location: in deep ocean water areas (far from the coast), intermediate areas or shallow water areas (close to the shore). This classification characterizes wave motions with different parameters. In deep waters, the free surface defined by the interface between air and water is generally subjected to a large oscillatory behavior, whereas in shallow waters waves break near the shore. Representing the visual complexity of these phenomena is a challenge, and the last 30 years have seen computer graphics evolve in order to address this issue.

Different models can be used to represent ocean dynamics: parametric description, spectral description as well as models from Computational Fluid Dynamics (CFD) and more specifically Navier-Stokes Equations (NSE). The first category aims at computing the path of water particles and describes the free surface with parametric equations based on real observations, obtained from buoys or satellite measurements \cite{Biesel}. The second category approximates the state of the sea by using waves spectrum \cite{PM64,He73,TMA} and computes waves distribution according to their amplitudes and frequencies. Finally, NSE can represent dynamics of all types of fluid, including the dynamical behavior of the ocean.

Ocean simulation methods in the computer graphics domain can therefore be classified into two main categories: \color{black}parametric/spectral methods that use oceanographic models, and physically-based methods relying on NSE. Parametric/spectral \color{black} models work best in deep waters where they accurately represent the periodical motion of the sea. But since they do not take into account the interactions with the bottom of the sea in shallow waters, only \color{black} physically-based methods deriving approximate solutions from NSE \color{black} can reproduce the complexity of ocean dynamics near the shore.

Another important characteristics of oceans for computer graphics is their complex optical properties. For example, the color of ocean waters, varying from green to deep blue, is related to the concentration of phytoplankton particles. Several other phenomena, such as foam, sprays, water properties (turbidity, bubbles, \ldots) and light-water exchanges must also be simulated at the rendering stage. Addressing the simulation of these phenomena is a precondition to obtain realistic ocean scenes.

This paper presents a survey of research works in ocean simulation and rendering in computer graphics. This includes different methods specifically designed for ocean scenes, but also more general water simulation techniques that can be applied to ocean simulation. Papers presenting specific methods for fluid simulation, intended for example for rivers \cite{Thon01UK,YNBH09} or fountains \cite{DBLPWan} are not covered here; interested readers can also refer to \cite{Iglesias04} where a more general survey of water simulation techniques is presented.

In sections \ref{section:simulation1} and \ref{section:simulation2}, we will focus on the methods specifically intended to model and animate the detailed surface of the ocean. We will use the classification presented in the previous paragraph: \color{black} parametric/spectral methods describing the surface using models from oceanography, and NSE-based methods that simulate the dynamic behaviour of ocean waves. Section \ref{section:rendu} is dedicated to realistic ocean rendering,  \color{black} particularly the representation of foam and sprays and the simulation of different light-water interactions. Finally, we will conclude by presenting possible perspectives for these techniques.


\section{Ocean dynamics simulation in deep water}
\label{section:simulation1}

\color{black}The methods presented in this section \color{black} use theoretical models and/or experimental observations to describe the ocean surface in deep water and represent swell effects. We can subdivide these methods into three main categories: first, those describing the ocean surface directly in the spatial domain, then those describing the surface in spectral domain and finally hybrid methods combining the two. As we will show in the following section, spatial methods use a height map computed as a sum of periodical functions, and animated with a simple phase difference, in order to represent the ocean surface. Spectral approaches use a wave spectrum to describe the surface in the spectral domain and a Fourier Transform to obtain its transformation in the spatial domain. Finally the combination of the two, called hybrid methods, produces convincing surfaces that can be animated easily.  

\subsection{Spatial domain approaches}
      \label{subsubsection:spatial}

      \begin{figure}
      \begin{center}
      \includegraphics[width=6cm]{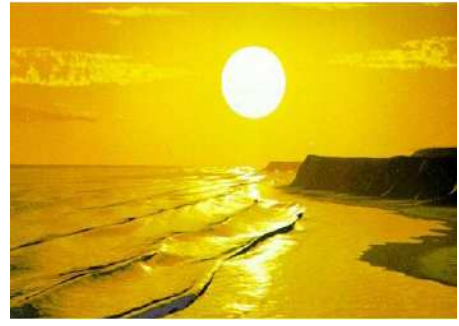}
      \end{center}
      \caption{Ocean surface obtained in \cite{Fournier}}
      \label{fig:fournier}
      \end{figure}

The main goal of spatial domain approaches is to represent the geometry of the water surface using a sum of periodical functions evolving temporally using a phase difference.

\color{black}\subsubsection{Early works}\color{black}

\begin{figure}
      \begin{center}
      \includegraphics[width=7cm]{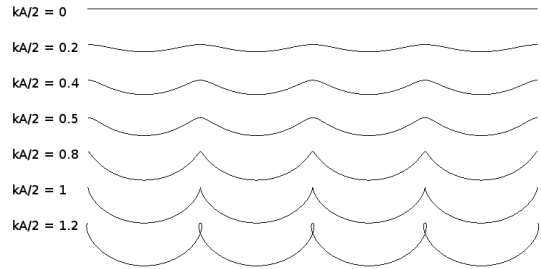}
      \end{center}
      \caption{Shapes of waves obtained using Eq. \ref{eq:gestner}}
      \label{fig:gestner}
      \end{figure}

This idea to combine series of sinusoids with high and low amplitudes was first proposed by Max \cite{Max81}. Ocean surface is represented as a height map with height $y=h(x,z,t)$ computed at each point ($x,z$) at time $t$ by:
\begin{equation}
 h(x,z,t) = -y_{0} + \sum_{i=1}^{N_w}A_{i}cos(k_{i_{x}}x+k_{i_{z}}z-w_{i}t)
\label{eq:gestner}
\end{equation}
where $N_w$ is the total number of waves, $A_{i}$ is the amplitude of the $i$-th wave, $\vec{k_{i}} = (k_{i_{x}}, k_{i_{z}})$ its wave vector, $w_{i}$ its pulsation and $y_{0}$ is the height of the free surface. The x-axis is oriented horizontally and points towards the coastline, the y-axis is vertical, and the z-axis is horizontal and aligned with the coastline. For each wave, the shape of the curve defined by the motion of a single point depends directly on the product between the amplitude $A_i$ and the wave number $k_i=||\vec{k_{i}}||$. 
\color{black}
If $k_{i}A_{i}<0.5$, this path is similar to a trochoïd. If $k_{i}A_{i}=0.5$, the shape is a cycloid. In all other cases ($k_{i}A_{i}>0.5$), this path cannot represent a realistic motion (see Figure \ref{fig:gestner}).
\color{black}

In order to obtain a realistic effect, the wave vector $\vec{k_{i}}$ of each wave is computed using scattering relationship in deep water domain, {\textit i.e.} supposing that the bottom of the sea is located at infinite depth:
\begin{equation}
 k_{i}=2\pi/\sqrt{\frac{gL_{i}}{2\pi}}
\end{equation}
with $g$ the gravitation constant and $L_{i}$ the wavelength of each individual wave.

The same idea was developed with a simple bump mapping approach, where normal vectors on a planar mesh are transformed using a sum of $20$ cycloids \cite{schachter}.
However, assuming that the bottom of the sea is at infinite depth limits the use of these methods since they don't include more complex phenomena such as breaking waves or waves refraction near the shore. Peachey \cite{Peachey} introduces a depth parameter to compute the wave vector $\vec{k_{i}}$ of each wave, using Airy wave theory:
\begin{equation}
k_{i}=2\pi/\sqrt{\frac{gL_{i}}{2\pi}tanh{\frac{2\pi d}{L_{i}}}}
\end{equation}
where $d$ is the depth of a point related to the bottom of the sea.

Fournier and Reeves \cite{Fournier} suggested modifying Gerstner’s theory of waves, where water particles positions are given by:
\begin{eqnarray}\label{eq:gerstner}
       \left\{
       \begin{array}{l}
       x=x_{0}-Ae^{ky_{0}}\sin(kx_{0}-wt)\\
       y=y_{0}-Ae^{ky_{0}}\cos(kz_{0}-wt)
       \end{array}
       \right.
      \end{eqnarray}
with $x$ (respectively $y$) the horizontal (resp. vertical) coordinate of a water particle at time $t$, $x_{0}$ and $y_{0}$ its coordinates at rest, $A$ the wave amplitude, $k$ the wave number and $w$ the wave pulsation. Fournier and Reeves enhance this model by taking into account the transformation of the path of water particles following the topological changes of the sea bed, and by transforming their circular path into a more realistic elliptic motion. This method permits to control the waves' shape, more or less crested, through the use of different parameters, and therefore yields a more realistic result (see Figure \ref{fig:fournier}).

Gonzato and Le Saec \cite{Gonz99} modify this model when the starting point of a \color{black}plunging breaking wave is detected near the
shoreline (\textit{i.e.} if the wave's crest starts to curl over). \color{black} This phenomenological modification results in the addition of two local functions applied to the wave's shape: a \textit{stretch } function imitates Biesel law by progressively stretching the wave along its crest, and a \textit{plunging} function simulates gravity. 

Ts’o and Barsky \cite{Ts'o} suggested that the refraction of waves could be  represented by the principle of light refraction formulated by Descartes. Their approach, called \textit{wave tracing}, consists in generating a spline surface by casting rays from the skyline in a uniform 2D grid, progressing with Bresenham's algorithm. The deviation of a ray is computed according to the depth difference of a cell to the next. However, a major drawback with this method is the lack of details on the generated surface when rays diverge significantly from a straight line, since in that case the number of rays defining the surface is too low. Gonzato and Le Saec \cite{Gonz00} address this problem by generating new rays in undersampled areas, offering a better representation of the ocean surface around bays or islands for instance. This method called ``Dynamic Wave Tracing'' 
can handle wave reflexion and diffraction due to obstacles. This approach was also used by Gamito and Musgrave \cite{Gamito02} to extract a phase map used in a parametric model, which can be animated easily.
The height map representing the ocean surface can be rendered using enhanced ray tracing \cite{Gonz00} or sphere tracing \cite{DCG07} methods.

\color{black}\subsubsection{GPU implementations}\color{black}

Series of periodic functions are particularly well suited to GPU computations, and in the last few years research works describing real time implementations have emerged.

Chen \textit{et al} \cite{ChenLi07} use four sinusoids, computed in a vertex shader and transferred to a bump map in a pixel shader to simulate ripples. Salgado and Conci \cite{Sal07} describe a real-time implementation of Fournier's method \cite{Fournier} using a vertex shader.
Schneider and Westermann \cite{Schneider01} compute a displacement map using Perlin noise in a vertex shader on the GPU to obtain interactive results.
Isidoro \textit{et al} \cite{Isidoro02} use a precomputed mesh perturbed by four sinusoids with low frequencies in a vertex shader. Ripples are obtained by using a bump map combining multiple textures. When using a limited number of input functions, visually convincing surfaces are obtained in interactive time (see Figure \ref{fig:isidoro}). Cui \textit{et al} \cite{Cui04} follow the same approach combined with an adaptive tessellation of the surface using the method of Hinsinger \textit{et al} \cite{Hinsingert} (described in Section \ref{hybrids}).

Chou and Fu \cite{Chou07} described a new GPU implementation for ocean simulation in order to take into account various interactions between the ocean surface and fixed or dynamic objects. The aim of this approach is to combine a particle system with a height-field obtained with a sum of sinusoids computed at each vertex of the surface mesh. Particles move according to the motion of each vertex but also to hydrodynamical forces \cite{24,25} due to solid objects interacting in the scene.

\color{black}\subsubsection{Adaptive schemes}\color{black}

In order to limit computation time, adaptive schemes have been proposed to limit computations only in visible areas and/or to reduce the number of periodic functions used to represent the surface dynamically, depending on the distance to the viewer. This approach consists in discarding high frequencies in distant areas, also limiting aliasing effects such as moir\'e patterns visible on the horizon. The ``grid project'' concept proposed by Johanson \cite{Johanson} consists in projecting the pyramid of vision onto the height map representing the water surface. Distant areas are automatically sampled at low resolution, whereas high resolution is used near the viewer.

Finch \cite{Fin04} and Lee \textit{et al} \cite{Lee06A,Lee06B} propose a Level of Details (LOD) tessellation by adapting the resolution of the surface to the distance between the viewer and the horizontal plane; using the angle between the camera and the surface, vertices outside the pyramid of vision are also discarded before rendering stage. The implementation by Lee \textit{et al} provides a $40\%$ gain in computation time compared to a fixed-resolution approach, and outperforms the implementation of Johanson \cite{Johanson} (tests were conducted on a Pentium 4 at \color{black}3.0GHz \color{black} and a GeForce 6600 GPU).

\begin{figure}
      \begin{center}
      \includegraphics[width=7cm]{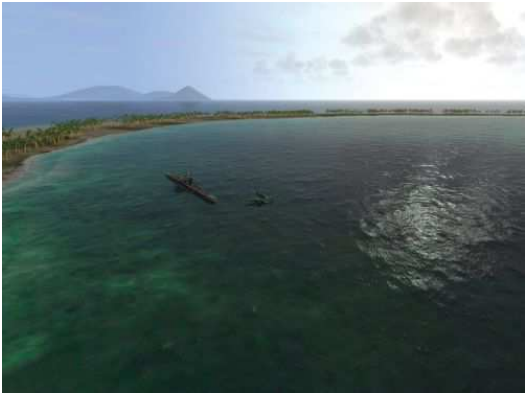}
      \end{center}
      \caption{Real-time ocean simulation from \cite{Krya05}}
      \label{fig:isidoro}
      \end{figure}

Kryachko \cite{Krya05} use \textit{vertex textures}, by computing two height maps representing the global motion of waves and fine scale details. The resolution is based on the distance to the viewer, allowing for fine-tune computations (see Figure \ref{fig:isidoro}).


\subsection{Fourier domain approaches} 
 \label{subsubsection:fourier}

This type of representation consists in describing the ocean surface using a spectral distribution of waves, obtained from theoretical or measured data. The spatial representation is then computed using \color{black}the most significant frequency components and an inverse fast Fourier transform (IFFT)\color{black}, 
giving a geometric characterization of the surface as a dynamic height map defined for each point $X$ by:
\begin{equation}
 h(X,t)=\sum_{i=1}^{N_s}\tilde{h}(\vec{k},t)e^{i\vec{k}X}
\end{equation}
where $N_s$ is the number of spectral components, \color{black}$\vec{k}$ \color{black} is the wave vector and $\tilde{h}(\vec{k},t)$ is the amplitude of the Fourier component obtained from a theoretic wave spectrum.

\color{black}\subsubsection{General methods}\color{black}

This idea was proposed by Mastin \textit{et al} \cite{Mastin} with the Pierson-Moskowitz spectrum \cite{PM64}. The amplitude of each Fourier component at time $t=0$ is given by:
     \begin{equation}
     \tilde{h_{0}}(\vec{k})=\frac{\alpha g^{2}}{16\pi^{4}f^{5}}e^{-\frac{5}{4}(\frac{f_{m}}{f})^{4}}
     \end{equation}
     where $\alpha=0.0081$ is called Phillips' constant, $g$ is the gravitation constant. The frequency peak $f_{m}$ is defined by:
     \begin{equation}
     f_{m}=\frac{0.13 g}{U_{10}}
     \end{equation}
     with $U_{10}$ the wind speed measured at $10$ meters above the surface.

Premoze and Ashikhmin \cite{Pre01} follow the same idea by using the JONSWAP spectrum \cite{He73}, a modified version of Pierson-Moskowitz's which reduces frequencies and thus raises waves' amplitudes. The amplitude of each Fourier component at time $t=0$ in that case is slightly different: 
     \begin{equation}
     \tilde{h_{0}}(\vec{k})=\frac{\alpha g^{2}}{16\pi^{4}f^{5}}e^{-\frac{5}{4}(\frac{f_{m}}{f})^{4}}e^{\ln(\gamma)e^{\frac{-(f-f_{m})^{2}}{2\sigma^{2}f_{m}^{2}}}}
     \end{equation}
     with:
     \begin{eqnarray*}&\sigma=&
     \left\{
      \begin{array}{l}
          0.07  \mbox{ if }f\leqslant f_{m}\\
          0.09  \mbox{ if }f>f_{m}
      \end{array}
     \right.
     \end{eqnarray*}

\begin{figure} 
      \begin{center}
      \includegraphics[width=7cm]{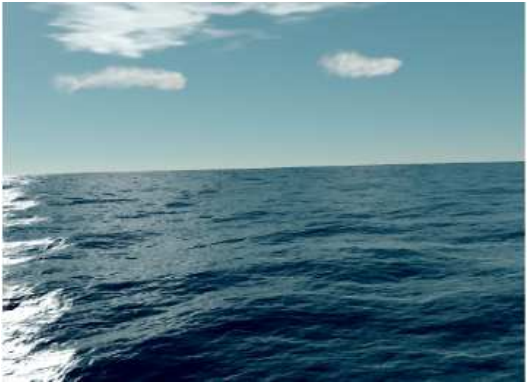}
      \end{center}
      \caption{Ocean surface obtained in \cite{Tess01}}
      \label{fig::Tessendorf}
      \end{figure}

The method presented by Tessendorf \cite{Tess01} was notably used in famous computer-generated scenes for \textit{Waterworld} and \textit{Titanic} productions. Again, this approach uses spectral components to describe the surface, computed by a Gaussian pseudo-random  generator and a theoretic wave spectrum due to Phillips: 
\begin{equation}
 \tilde{h_{0}}(\vec{k})=\frac{1}{\sqrt{2}}(\gamma_{r}+i\gamma_{i})\sqrt{P_{h}(\vec{k})}
\end{equation}
with $\gamma_{r}$ and $\gamma_{i}$ defined as constants. The spectrum $P_{h}(\vec{k})$ is given by:
\begin{equation}
 P_{h}(\vec{k})=C\frac{e^{-1/(kL)^{2}}}{k^{4}}|\vec{k}.\vec{w}|^{2}
\end{equation}
where $C$ is a constant, $\vec{k}$ is the wave vector, $k$ is the wave number, $\vec{w}$ is the wind direction, $V$ is its speed and $L=\frac{V^{2}}{g}$.
The choice of a random generator is justified by the Gaussian distribution of waves often observed in deep sea areas. In order to obtain fine details on the surface, each component is slightly perturbed by a noise function, producing more or less crested waves at any resolution (see Figure \ref{fig::Tessendorf}). Cieutat \textit{et al} \cite{CGG03} enhance this model in order to take in account water waves forces applied to a ship, implemented in a ship training simulator.

Although results obtained with this approach seem visually convincing for a distant viewer, when zooming towards the surface, waves' shapes tend to look unrealistic since the resolution of the heightmap is fixed. In that case it becomes necessary to generate a new surface using more spectral components. This problem is addressed by interactive methods proposed recently.

\color{black}\subsubsection{Level-Of-Detail and GPU implementations}\color{black}

Hu \textit{et al} \cite{Hu06} apply a LOD approach to Tessendorf's method by representing the ocean with two surfaces. The first one, with a high, fixed resolution, is animated using a displacement map stored in a vertex shader, while the second one is sampled adaptively and applied as a bump map stored in a pixel shader. This approach produces a large ocean surface in real-time by animating only visible parts (at approx. $100$ frames/sec. on GeForce 3 GPU card).
Mitchell \cite{Mitchell} implemented Tessendorf's method on GPU by only considering frequencies generating a significant perturbation of the surface. A white noise is first generated using Phillips' spectrum, then frequencies are divided in two sets. \textit{Low }frequencies accounting for the global motion of waves are stored as a displacement map in a vertex shader. \textit{High} frequencies, representing fine details, are stored in a normal map used for rendering. Since only a part of the spectrum is animated, a realistic ocean surface is obtained at interactive rates; another implementation by Chiu and Chang \cite{Chiu06} combine this approach with an adaptive tessellation method \cite{Johanson} (see previous section).

An adaptive spectrum sampling is presented by Frechot \cite{Fre06} in order to obtain a more realistic surface with capillary waves or riddles. The main idea of this approach is to re-sample the spectrum corresponding to high frequencies areas, depending on the distance to the viewer. Robine and Frechot \cite{Fre06.2} also described a real-time additive sound synthesis method applied to the ocean surface. Sound synthesis methods are commonly used to efficiently sum a large number of sinusoidal components called partials. Here, ocean waves are considered as partials by shifting from the frequency and time domains to the wave number and spatial ones.

\subsection{Hybrid approaches}
          \label{hybrids}

\begin{figure} 
      \begin{center}
      \includegraphics[width=7cm]{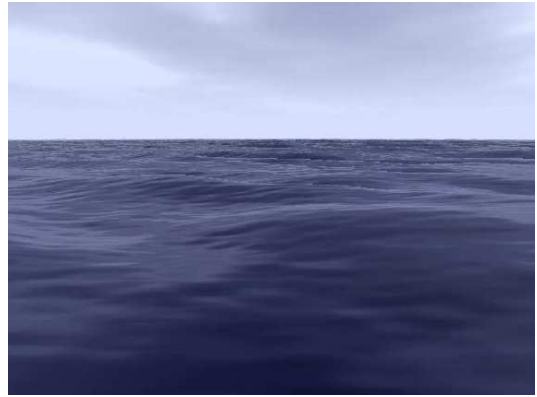}
      \end{center}
      \caption{Ocean surface obtained in \cite{TDG00}}
      \label{fig::Thon}
      \end{figure}

Hybrid approaches are a combination of spatial and spectral definitions, generating a geometric representation of the surface while describing the components of wave trains in a realistic manner.

The method proposed by Thon \textit{et al} \cite{TDG00,Thon02} shows a combination of a linear superposition of trochoids, which characteristics are synthesized using Pierson and Moskowitz's spectrum. In order to get fine scale details, a 3D turbulence function is added to the surface; this method provides very realistic results, as shown on Figure \ref{fig::Thon}. 

Lee \textit{et al.} \cite{Lee07} combine a superposition of sinusoids and the TMA spectrum (Texel, Marson and Arsole, based on Jonswap's) \cite{TMA}, creating a better statistical distribution of waves. It also offers better control for the end-user who can change parameters such as the bottom depth, thus enabling him to generate different types of water volumes (lakes, rivers, etc).

Xin \textit{et al} \cite{XinFS06}  extend the method of \cite{Thon02} to simulate the effects of wind variations over the surface by modulating each spectrum frequency accordingly. 

Hinsingert \textit{et al} \cite{Hinsingert} apply a LOD scheme to the method of \cite{TDG00} by reducing both the sampling resolution of the surface mesh and the number of trochoidal components depending on the distance to the viewer. This approach reduces the amount of computations for the ocean simulation and tends towards real-time rates (approximately $20$ to $30$ frames per sec. on a GeForce 2 GPU), by using only a subset of the components.

A real-time simulation method for ocean scenes was proposed by Lachman \cite{Lach07} with a framework representing the surface either in the spatial domain by combining a spectrum and a sum of sinusoids or by using the approach of \cite{Tess01}. This method is highly controllable, allowing the user to modify parameters such as the bottom depth or the agitation of the sea (defined by the Beaufort Scale). A LOD representation of the surface is also implemented to reduce computation costs. The main idea consists in defining several resolution levels from the projection of the pyramid of vision on the surface, and meshing the surface accordingly.

\color{black}
\subsection{Discussion}

\begin{table*}
\begin{center}
\begin{tabular}[t]{|l|c|c|c|c|c|c|}
\hline
\multicolumn{2}{|c|}{} & \multicolumn{2}{c|}{Model} & \multicolumn{3}{c|}{Implementation}\\  \cline{3-7}
\multicolumn{2}{|c|}{}  &Spatial &Spectral &Wave refraction	&GPU &Level-Of-Details	\\ 
  \hline
1980	&Schachter 	\cite{schachter}	&x 	&	&	&	&	\\
  \hline
1981	&Max 	\cite{Max81}	&x 	&	&	&	&	\\
  \hline
1986	&Fournier et al. 	\cite{Fournier}	&x 	&	&	&	&	\\
  \hline
1987	&Mastin et al. 	\cite{Mastin}	&	&x 	&	&	&	\\
  \hline
1987	&Ts'o et al. 	\cite{Ts'o}	&x 	&	&x 	&	&	\\
  \hline
1999	&Gonzato et al. 	\cite{Gonz99}	&x 	&	&	&	&	\\
  \hline
2000	&Gonzato et al. 	\cite{Gonz00}	&x 	&	&x 	&	&	\\
  \hline
2000	&Thon et al.	\cite{TDG00}	&x 	&x 	&	&	&	\\
  \hline
2001	&Premoze et al. 	\cite{Pre01}	&	&x 	&	&	&	\\
  \hline
2001	&Tessendorf 	\cite{Tess01}	&	&x 	&	&	&	\\
  \hline
2001	&Schneider et al. 	\cite{Schneider01}	&x 	&	&	&x 	&	\\
  \hline
2002	&Gamito et al.	\cite{Gamito02}	&x	&	&x 	&	&	\\
  \hline
2002	&Hinsinger et al. 	\cite{Hinsingert}	&x 	&x 	&	&x 	&x	\\
  \hline
2002	&Isidoro et al. 	\cite{Isidoro02}	&x 	&	&	&x 	&	\\
  \hline
2002	&Thon et al. 	\cite{Thon02}	&x 	&x 	&	&	&	\\
  \hline
2003	&Cieutat et al. 	\cite{CGG03}	&	&x 	&x 	&	&	\\
  \hline
2004	&Finch	\cite{Fin04}	&x 	&	&	&x 	&x	\\
  \hline
2004	&Johanson 	\cite{Johanson}	&x 	&	&	&x 	&x	\\
  \hline
2005	&Kryachko	\cite{Krya05}	&x 	&	&	&x 	&x	\\
  \hline
2005	&Mitchell 	\cite{Mitchell}	&	&x 	&	&x 	&	\\
  \hline
2006	&Chiu et al.	\cite{Chiu06}	&	&x 	&	&x 	&x	\\
  \hline
2006	&Frechot 	\cite{Fre06}	&	&x 	&	&x 	&x	\\
  \hline
2006	&Hu et al.	\cite{Hu06}	&	&x 	&	&x 	&x	\\
  \hline
2006	&Lee et al. 	\cite{Lee06A}	&x 	&	&	&x 	&x	\\
  \hline
2006	&Lee et al. 	\cite{Lee06B}	&x 	&	&	&x 	&x	\\
  \hline
2006	&Robine et al.	\cite{Fre06.2}	&	&x 	&	&x 	&	\\
  \hline
2006	&Xin et al.	\cite{XinFS06}	&x 	&x 	&	&x	&x	\\
  \hline
2007	&Chen et al. 	\cite{ChenLi07}	&x	&	&	&x 	&	\\
  \hline
2007	&Chou et al. 	\cite{Chou07}	&x 	&	&x 	&x 	&	\\
  \hline
2007	&Lachman	\cite{Lach07}	&x 	&x 	&	&x 	&x	\\
  \hline
2007	&Lee et al. 	\cite{Lee07}	&x 	&x 	&	&x 	&	\\
  \hline
2007	&Salgado et al. 	\cite{Sal07}	&x 	&	&	&x 	&	\\
  \hline
\end{tabular}
\end{center}
\caption{Classification of the simulation methods for deep water presented  in section \ref{section:simulation1}}
\label{fig:class1}
\end{table*}

The main advantage of spatial domain approaches is their ability to produce a simple and fast simulation of the ocean surface, but they require a large number of periodical functions for visually plausible results. Several optimizations have been proposed such as GPU evaluation or adaptive schemes to reduce this number according to the distance to the viewer. However, using sine or cosine functions induces a too rounded shape of waves and the surface appears too smooth.\\
Spectral domain methods address this problem by using oceanographic data directly and allow to disturb the surface according to physical parameters such as wind speed, which brings more realistic results. Unfortunately these methods lack from global control of ocean waves.\\
In this context, hybrid approaches represent a good compromise between the visually plausible results obtained by spectral approaches and the global control offered by spatial domain methods, and allow to obtain fast and effective simulations of the ocean surface's dynamic behaviour. A complete summary of the methods presented in this section is shown on Table \ref{fig:class1}.

However, all these methods only consider deep water phenomena, in which the ocean surface is being subjected to small perturbations. Indeed, in shallow water parametric or spectral approaches cannot faithfully reproduce ocean dynamics near coasts, \textit{e.g.} breaking waves.  To address this complexity, we have to focus on approaches that consider interactions and collisions between the ocean and the shore, which are presented in the next section.

\section{Ocean dynamics simulation in shallow water}
\label{section:simulation2}

Navier-Stokes equations (NSE) are able to capture the dynamic complexity of a fluid flow. Incompressible flow is usually considered for large volumes of fluid like oceans:
\begin{eqnarray}
        \label{eq:incomp}
        &\nabla \vec{U} = 0\\
       \label{eq:mouvement}
       &\frac{\partial \vec{U}}{\partial t}+ \vec{U}\nabla\vec{U}+ \frac{\vec{\nabla} p}{\rho}-\mu\frac{\nabla^{2}\vec{U}}{\rho}-\vec{g} = 0  
\end{eqnarray}
\color{black}
with $\vec{U}=(u,v,w)$ the velocity of the fluid, $\mu$ its viscosity, $p$ its pressure, $\rho$ its density, and $\vec{g}$ representing gravity $(0,9.81,0)$. Operator $\nabla\vec{U}$ (resp. $\nabla^{2}\vec{U}$) is the gradient (resp. Laplacian) of $\vec{U}$ and is defined by $\nabla\vec{U}=(\frac{\partial u}{\partial t},\frac{\partial v}{\partial t},\frac{\partial w}{\partial t})$ (resp. $\nabla^{2}\vec{U}=\frac{\partial^{2} u}{\partial^{2} t}+\frac{\partial^{2} v}{\partial^{2} t}+\frac{\partial^{2} w}{\partial^{2} t}$).             
The first equation guarantees mass conservation, \textit{i.e.} the density of the fluid remains constant over time. The second one guarantees moment conservation: the acceleration $\frac{\partial \vec{U}}{\partial t}$ of the fluid is equal to the sum of the applied forces weighted by its density. 
Other formulations for fluid behavior include shallow water equations (or Saint-Venant equations), preferably used whenever the horizontal scale is greater than the vertical one.

There are two main ways of discretizing these equations. On the one hand, Eulerian approaches use a 2D or a 3D grid where the fluid goes from cell to cell. On the other hand, Lagrangian approaches are based on particles carrying the fluid, thus representing small fluid volumes. Hybrid techniques were also developed to combine these two techniques, adding small-scale details to Eulerian simulations such as bubbles or spray. An excellent tutorial on physically-based simulation of fluids for Computer Graphics is presented by Bridson \textit{et al} \cite{Bridson06}. Adabala and Manohar \cite{Adabal02} proposed a survey of most physically-based fluid simulation techniques.
Since numerous phenomena can be simulated by these methods (fire, smoke, lava, etc.) we will focus on liquids and more specifically on ocean waters.

\subsection{Eulerian approaches}

Kass and Miller \cite{Kass90} were the first to use a physically-based realistic approach to represent the ocean surface. This method consists in solving Saint-Venant equations in 2D in order to obtain a heightmap. This was later extended by Chen and Lobo \cite{Chen95} to solve NSE taking into account pressure effects, hence generating a more realistic surface modulation.
           
A full 3D resolution of NSE was introduced in the Computer Graphics community by Foster and Metaxas \cite{Foster96,Foster97}, inspired from a physical approach proposed by Harlow and Welch \cite{Harlow65} based on a uniform voxel decomposition of the 3D space. Velocity of the fluid is defined at the centers of each face of voxels; gradient and Laplacian are computed using finite differences between faces of adjacent voxels. Divergence at a given voxel is obtained from the updated velocities at its faces. In order to guarantee fluid incompressibility, the pressure field at a given voxel is modified according to the divergence. 
Virtual particles are placed in the voxel grid to extract the fluid surface, \textit{i.e.} to discover its location, however the resulting surface does not look visually convincing. This problem was later addressed by Foster and Fedkiw \cite{Foster01} who represent it as an implicit, \textit{level-set} surface evolving according to the fluid's velocity. Nevertheless, visual results show compressibility artifacts since a noticeable volume of fluid disappears during the animation.

In order to solve this problem, Enright \textit{et al} \cite{Enr02} place particles above and below the surface, and advect them according to the velocity of the voxel they belong to. The surface is then computed by interpolating between the two implicit functions obtained from the locations of these particles. Since this approach called \textit{particles level-set method} can be applied to all types of fluids, it became very popular and was widely used for many purposes, including the simulation of interactions between fluids and solids \cite{Carlson04, Wang_05} or between different types of fluids \cite{Losa06}. The main drawback is high computational costs, which can be reduced by using an adaptive, octree-like structure \cite{Losa04} to refine the voxel grid only near the fluid-solid or fluid-fluid interface.

\begin{figure} 
          \begin{center}
          \includegraphics[width=6cm]{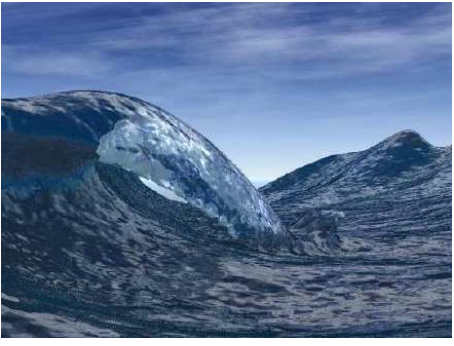}
          \end{center}
          \caption{Eulerian breaking waves simulation from \cite{Milh04}}
          \label{fig::euler}
          \end{figure}

This approach permits to capture the dynamic behavior of fluid surfaces in a realistic way and can be used for oceanic scenes.
Enright \textit{et al} \cite{Enr02} simulate breaking waves by constraining the velocities in the grid using parametric equations \cite{RO98}. Although realistic results are obtained, this approach can only handle limited types of waves (spilling or \textit{rolling} waves). 
Mihalef \textit{et al} \cite{Milh04} simulate breaking waves by constraining velocities inside the grid using the parametric approach of Thon \textit{et al} \cite{TDG00} in 2D. The initial horizontal and vertical velocities of each element of fluid are given by: 
\begin{eqnarray*}
           \left\{
           \begin{array}{l}
           u = Awe^{k_{z}}\cos(k_{x})\\
           v = Awe^{k_{z}}\sin(k_{x})\\
           \end{array}
           \right.
           \end{eqnarray*}
where, as in \cite{TDG00}, $A$ is the amplitude of a wave, $\vec{k} = (k_{{x}}, k_{{z}})$ its characteristic vector and $w$ its pulsation.
The user can obtain any type of shore-break by choosing from a library of precomputed profiles. The full 3D simulation is then computed by extruding the desired profile along the parallel direction to the shore (see Figure \ref{fig::euler}).

Th\"{u}rey \textit{et al} \cite{Thuerey07} simulate breaking waves in real-time, by solving Saint-Venant equations in 2D. A breaking wave is obtained by detecting regions with high curvatures on the heightmap representing the fluid, and adding a new mesh to account for the water falling from the crest. Each vertex of this mesh is projected in the 2D grid and advected using the corresponding cell.


\subsection{Lagrangian approaches}
 \label{subsubsection:lagrangian}
In Lagrangian approaches, the fluid is represented by a set of particles following physical laws. Particle systems were first introduced in the computer graphics community by Reeves \cite{Reeves} to simulate natural phenomena such as water, fire, clouds or smoke.
Miller and Pearce \cite{MP89} simulate interactions between particles by connecting them with the aid of springs to represent attraction or repulsion forces between neighboring particles, hence simulating viscous liquid flows. Terzopoulos \textit{et al} \cite{Terzo89} later extended this method to simulate morphing from a solid to a liquid by modifying the stiffness of each spring. Tonnesen \cite{Tonn91} simulates the same kind of effects by solving heat transfer equations on the particles set.

Although realistic results are obtained with these methods, they are unable to simulate the inner dynamic complexity of fluids described by NSE. Stam and Fiume \cite{Stam95} introduced the \textit{Smoothed Particle Hydrodynamics} (SPH) model in computer graphics, originally used in astrophysics to study the dynamics of celestial objects as well as their interactions \cite{Lucy77}, and formalized by Monaghan \cite{Mon94} for fluids. In that case, NSE are simplified and linearized. By considering that each particle has its own mass and that their sum gives the overall mass of the fluid, Equation \ref{eq:incomp} can be omitted since the overall mass remains constant over time. In equation \ref{eq:mouvement}, the non-linear advection term, representing the fluid's transport related to its velocity in the Eulerian case, can be omitted: the fluid is transported by the particles themselves. Finally only one linear equation is necessary to describe the fluid's velocity $\vec{U}$:
\begin{equation}\label{eq::stokes_lagrangien}
\frac{\partial{\vec{U}}}{\partial{t}}=\frac{1}{\rho}(-\nabla p+\mu \nabla^{2}\vec{U} + \rho g)
\end{equation}
Let $\vec{f}=-\nabla p+\mu \nabla^{2}\vec{U} + \rho G$ represent the total amount of forces applied on a particle. The term $\nabla p$ accounts for pressure forces, $\mu \nabla^{2}\vec{U}$ for viscosity forces, and $\rho G$ for exterior forces. The acceleration $a$ of a particle is then given by:
\begin{equation}\label{eq:acceleration}
a = \frac{\partial{\vec{U}}}{\partial{t}}=\frac{1}{\rho}(\vec{f}^{press}+\vec{f}^{visc}+\vec{f}^{ext})
\end{equation}
where $\vec{f}^{press}$ (resp. $\vec{f}^{visc}$ and $\vec{f}^{ext}$) represents pressure (resp. viscosity and exterior forces).
The main idea in the SPH approach consists in defining particles as potential implicit surfaces (also known as \textit{blobs} or \textit{metaballs}) influencing their neighbors. A particle represents an estimation of the fluid in a given region rather than a single molecule. \color{black}Force computations \color{black} rely on the interpolation of any scalar value $S_i$ for a given particle $i$, depending on its neighbors:
\begin{equation}\label{eq::Ai}
S_{i}=\sum_{j=1}^{N_p}m_{j}\frac{S_{j}}{\rho_{j}}W(r,R)
\end{equation}
where $N_p$ is the total number of particles, $m_{j}$ (resp. $S_{j}$) is the mass (resp. the scalar value) of particle $j$, $r$ is the distance between particles $i$ and $j$, $R$ is the maximal interaction radius, and $W$ is a symmetric interpolation function. The gradient $\nabla S_{i}$ and Laplacian $\nabla^{2} S_{i}$ are given by:
\begin{equation}\label{eq::grad_Ai}
\nabla S_{i}=\sum_{j=1}^{N}m_{j}\frac{S_{j}}{\rho_{j}}\nabla W(r,R)
\end{equation}
\begin{equation} \label{eq::lapl_Ai}
\nabla^{2} S_{i}=\sum_{j=1}^{N}m_{j}\frac{S_{j}}{\rho_{j}}\nabla^{2} W(r,R)
\end{equation}
Equation \ref{eq::Ai} can then be used to compute the density of a particle $i$; pressure and viscosity computation will rely on equations \ref{eq::grad_Ai} and \ref{eq::lapl_Ai}.

The SPH model was extended by several authors \cite{Desbrun96, SAC99} to represent elastic deformations, lava flows, etc. by simulating heat transfer and varying viscosity. M\"{u}ller \textit{et al} \cite{Muller03} presented an extension to simulate low compressible liquids such as water. 
Clavet \textit{et al} \cite{Clavet} represent visco-elastic fluids by coupling particles with a mass-spring system; particles advection is governed by Navier-Stokes equations but also Newtonian spring dynamics, which slows the fluid down depending on springs stiffness.
Another extension by M\"{u}ller \textit{et al} \cite{Muller05} represent interactions between fluids with different phases by modifying the viscosity between neighboring particles. This method can also simulate bubbles generated by heating, using varying densities for fluid particles to represent the amount of gasification.

However, two problems are inherent to the SPH approach. First of all, compressibility effects can be observed when the volume does not remain constant over time, yielding non-realistic visualizations. Besides, a huge amount of particles is needed to simulate large volumes of water with fine-scale details, leading to high computation and memory consumption. These problems were addressed by several authors.

Premoze \textit{et al} \cite{Pre03} introduced the \textit{Moving Particle Semi-implicit} method (MPS) in computer graphics from physically-based models \cite{Kosh96, YKO96}. This approach is an extension of SPH where the density of a particle is modified to guarantee constant pressure through the solving of a Poisson equation. It was extended by Wang \textit{et al}  \cite{Wang06} to simulate 2D breaking waves, combined to obtain a 3D surface. 
Becker and Teschner \cite{Becker} guarantee near-incompressibility by modifying pressure depending on density variations. Since pressure is directly related to attraction or repulsion forces between particles, it is possible to ensure a near constant distance between neighboring particles, hence a near constant volume. An alternative method was also presented more recently by Solenthaler and Pajarola \cite{SolenthalerP09}.

Adaptive schemes were proposed to reduce the number of particles and limit computation and memory costs for simulating large volumes of fluid.
Desbrun and Cani \cite{DC99} merge or split particles according to their density. The overall mass of the system is kept constant by re-computing \color{black}particle masses \color{black} as soon as they are subjected to one of these operations. In order to guarantee symmetric interactions between particles with different masses, a \textit{shooting / gathering} scheme ensures that attraction or repulsion forces are equal. Although this approach reduces computation costs, visual artifacts can be observed during animations because of instantaneous merging or splitting of particles near the surface. The adaptive scheme proposed by Hong \textit{et al} \cite{HHK07} consists in varying the size and number of particles according to their position relative to a fixed set of \textit{layers} defined by the user in a preprocessing step. Finally, Adams \textit{et al} \cite{Adams} obtain adaptively sized particles by defining merging or splitting processes, depending on the distance to the viewer or to the surface. This ensures that visual artifacts are avoided near the surface since only particles deep inside the fluid can be modified. This approach was implemented on GPU by Yan \textit{et al} \cite{1568695} to simulate breaking waves in real-time.


\subsection{Hybrid approaches}
 \label{subsubsection:physical_hybrids}

      \begin{figure} 
      \begin{center}
      \includegraphics[width=6cm]{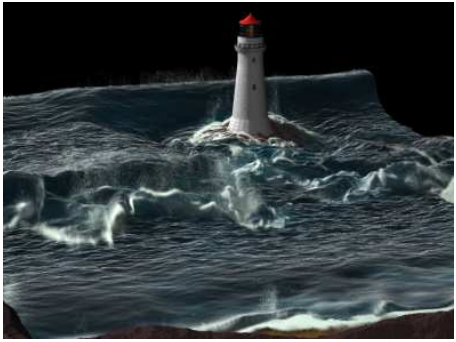}
      \end{center}
      \caption{Hybrid simulation from \cite{Losa07}}
      \label{fig::semi_lagrang}
      \end{figure}
The main idea in hybrid approaches is to simulate the main body of fluid using an Eulerian method, and fine-scale details such as foam, spray or bubbles with a Lagrangian method. By adding small details, very realistic results are obtained, and interest for this approach is growing more and more with the development of efficient simulations.
{O'B}rien and Hodgins \cite{Obrien95} couple an Eulerian method with a particle system to represent sprays. Navier-Stokes equations are solved in 2D, and sprays are generated depending on the magnitude of the vertical velocity at each cell; these particles are then advected using the velocity in the grid.
The Eulerian simulation of Takahashi \textit{et al} \cite{Taka03} is coupled with particles generated in high curvature regions, then advected independently.
Greenwood and House \cite{Green04} use a level-set approach \cite{Enr02} coupled with particles generated in high curvature regions, advected using the velocity of their corresponding cell. The friction of the fluid is also taken into account to generate bubbles.
This method was later extended by Zheng \textit{et al} \cite{Zhen06} by deforming each bubble based on surface tension, thus avoiding non-realistic spherical bubbles.

Kim \textit{et al} \cite{Kim06} extend an adaptive level-set approach \cite{Losa04} to generate particles according to the temporal variation of the volume of fluid in each cell. The main idea consists in using particles advected below the surface, transformed into water or air at the rendering stage. Since the number of transformed particles depends on the volume of water lost in a cell, this characterizes the turbulence in that cell and hence generates particles in highly perturbed regions.\\

\color{black}
Yuksel et. al \cite{waveparticle} introduced the concept of \textit{wave particles} to simulate waves and their interactions with solid objects.
Wave particles are defined by a disturbance function characterized by an amplitude, a propagation angle and a maximal interaction distance. The fluid itself is represented by a height-field whose values are computed according to neighbouring wave particles. When waves collide with solid objects, propagation angles are re-computed according to the curvature at collision points and new particles are generated to obtain waves refraction. 

Th\"{u}rey \textit{et al} \cite{Thur06} proposed to couple 2D and 3D simulations using Saint-Venant equations solved with a Lattice-Boltzmann method \cite{Fdh87}. The 2D simulation accounts for global motion of the surface, but fine-scale details are represented in 3D. These two layers interact with each other to generate foam and sprays, by creating particles according to the velocity in each cell and surface tension at the air-water interface. This work was later extended with SPH to represent interactions between particles in \cite {Thur07}.
\color{black}
The same approach is followed by Losasso \textit{et al} \cite{Losa07} to combine an Eulerian method with SPH particles whose number is computed from the divergence of the fluid's velocity, thus simulating bubbles or sprays generated by breaking waves (see Figure \ref{fig::semi_lagrang}).


\subsection{Discussion}

\begin{table*}
\begin{center}
\begin{tabular}[t]{|l|c|c|c|c|c|c|c|}
  \hline
\multicolumn{2}{|c|}{} & \multicolumn{2}{c|}{Model} & \multicolumn{2}{c|}{Resolution} & \multicolumn{2}{c|}{Implementation}\\  \cline{3-8}
\multicolumn{2}{|c|}{} 	&Eulerian	&Lagrangian	&2D-based &Full 3D &Incompressibility	&LOD	\\
  \hline
1989	&Miller et al.	\cite{MP89}	&	&x 	&	&x	&	&	\\
  \hline
1989	&Terzopoulos et al.	\cite{Terzo89}	&	&x 	&	&x	&	&	\\
  \hline
1990	&Kass et al.	\cite{Kass90}	&x 	&	&x	&	&	&	\\
  \hline
1991	&Tonnesen	\cite{Tonn91}	&	&x 	&	&x	&	&	\\
  \hline
1995	&Stam et al.	\cite{Stam95}	&	&x 	&	&x	&	&	\\
  \hline
1995	&Chen et al.	\cite{Chen95}	&x 	&	&x	&	&	&	\\
  \hline
1995	&{O'B}rien et al.	\cite{Obrien95}	&x 	&x 	&x	&	&	&	\\
  \hline
1996	&Desbrun et al.	\cite{Desbrun96}	&	&x 	&	&x	&	&	\\
  \hline
1996	&Foster et al.	\cite{Foster96}	&x 	&	&	&x	&	&	\\
  \hline
1997	&Foster et al.	\cite{Foster97}	&x 	&	&	&x	&	&	\\
  \hline
1999	&Desbrun et al.	\cite{DC99}	&	&x 	&	&x	&	&x	\\
  \hline
1999	&Stora et al.	\cite{SAC99}	&	&x 	&	&x	&	&	\\
  \hline
2001	&Foster et al.	\cite{Foster01}	&x 	&	&	&x	&	&	\\
  \hline
2002	&Enright et al.	\cite{Enr02}	&x 	&	&	&x	&x	&	\\
  \hline
2003	&Premoze et al.	\cite{Pre03}	&	&x 	&	&x	&x	&	\\
  \hline
2003	&Muller et al.	\cite{Muller03}	&	&x 	&	&x	&	&	\\
  \hline
2003	&Takahashi et al.	\cite{Taka03}	&x 	&x 	&	&x	&x	&	\\
  \hline
2004	&Carlson et al.	\cite{Carlson04}	&x 	&	&	&x	&x	&	\\
  \hline
2004	&Greenwood et al.	\cite{Green04}	&x 	&x 	&	&x	&	&	\\
  \hline
2004	&Losasso et al.	\cite{Losa04}	&x 	&	&	&x	&x	&x	\\
  \hline
2004	&Mihalef et al.	\cite{Milh04}	&x 	&	&x	&	&x	&	\\
  \hline
2005	&Clavet et al.	\cite{Clavet}	&	&x 	&	&x	&	&	\\
  \hline
2005	&Muller et al.	\cite{Muller05}	&	&x 	&	&x	&	&	\\
  \hline
2005	&Wang et al.	\cite{Wang_05}	&x 	&	&	&x	&x	&	\\
  \hline
2006	&Kim et al.	\cite{Kim06}	&x 	&x 	&	&x	&	&x	\\
  \hline
2006	&Losasso et al.	\cite{Losa06}	&x 	&	&	&x	&x	&	\\
  \hline
2006	&Thurey et al.	\cite{Thur06}	&x 	&x 	&x	&x	&x	&	\\
  \hline
2006	&Wang et al.	\cite{Wang06}	&	&x 	&x	&x	&x	&	\\
  \hline
2006	&Zheng et al.	\cite{Zhen06}	&x 	&x 	&	&x	&x	&	\\
  \hline
2007	&Adams et al.	\cite{Adams}	&	&x 	&	&x	&	&x	\\
  \hline
2007	&Becker et al.	\cite{Becker}	&	&x 	&	&x	&x	&	\\
  \hline
2007	&Thurey et al.	\cite{Thuerey07}	&x 	&	&x	&x	&	&	\\
  \hline
2007	&Thurey et al.	\cite{Thur07}	&x 	&x 	&x	&	&x	&	\\
  \hline
2007	&Yuksel et al.	\cite{waveparticle}	&x 	&x 	&x	&x	&	&	\\
  \hline
2008	&Hong et al.	\cite{HHK07}	&	&x 	&	&x	&	&x	\\
  \hline
2008	&Losasso et al.	\cite{Losa07}	&x 	&x 	&	&x	&x	&	\\
  \hline
2009	&Solenthaler et al.	\cite{SolenthalerP09}	&	&x 	&	&x	&x	&	\\
  \hline
2009	&Yan et al.	\cite{1568695}	&	&x 	&	&x	&	&x	\\

  \hline
\end{tabular}
\end{center}
\caption{Classification of the simulation methods for shallow water presented  in section \ref{section:simulation2}}
\label{fig:class2}
\end{table*}

The main advantage of Eulerian approaches is their ability to simulate a large scope of different phenomena, explaining why they received much attention from the computer graphics community.  Nevertheless, for fine-scale details such as spray or bubbles, NSE need to be \color{black}finely discretized \color{black} which in turn demands high computations and memory consumption.  Another drawback is the use of fixed grids, forbidding the fluid to flow outside the grid.

On the other hand, Lagrangian approaches can be used to simulate a wide range of phenomena. Their main advantage is their ability to represent fine-scale details and to flow anywhere in a virtual environment, but a large number of particles is usually needed to obtain realistic results. This problem can be alleviated using adaptive split-and-merge schemes to reduce computation costs. However it is worth noticing that most of the computation time for one particle is spent in testing neighboring particles or other objects for collision. Therefore a broad-phase collision detection is usually implemented by storing particles in a virtual grid, meaning that Eulerian or Lagrangian approaches share common problems such as defining an appropriate size for the grid's cells. This is also illustrated by hybrid methods which produce realistic results by adding details to Eulerian approaches using particle systems. \color{black}The literature presented in section \ref{section:simulation2} is summarized on Table \ref{fig:class2}.\color{black}

\color{black}
\section{Realistic ocean surface rendering and lighting}
\label{section:rendu}

In the previous sections we presented different approaches to obtain a plausible \textit{shape} of the ocean surface, in deep water or near the shoreline. This surface can then be rendered using water shaders implementing light reflection and refraction described by Fresnel equations. Nevertheless, the ocean's visual aspect is characterized by numerous interactions. Their impact is visible through several phenomena such as foam, sprays and light interactions, addressed by several methods in computer graphics. Some of them were already mentioned earlier, when the fluid simulation algorithm includes these phenomena directly (\textit{e.g.} bubbles), However in most cases rendering is performed in a post-processing step. We focus on the following on different methods designed to enhance the realism of ocean scenes.
\color{black}
\subsection{Foam and spray} 
\label{ecume_embruns}
Foam and spray rendering methods can be divided into two types: on the one hand, methods based on the computation of foam amount at one point on the ocean according to empirical models, and on the other hand, methods based on particle systems and rendered as point clouds.

    \subsubsection{Empirical models}
\begin{figure} 
      \begin{center}
      \includegraphics[width=7cm]{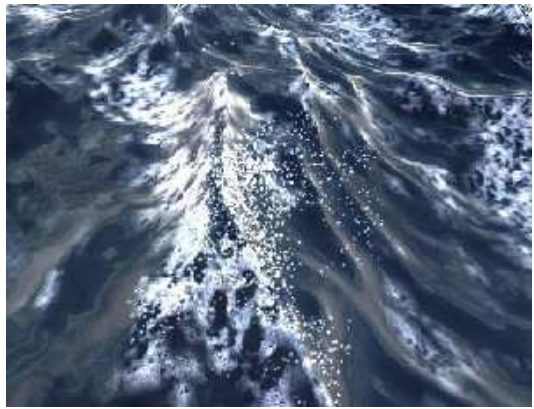}
      \end{center}
      \caption{Foam rendering from \cite{Jens01}}
      \label{fig:foam_pre_jens}
      \end{figure}

Jensen and Golias \cite{Jens01} compute the amount of foam at a given point according to its height: if the height difference with its neighbors is greater than a predefined threshold, foam is generated and rendered using a semi-transparent texture (see Figure \ref{fig:foam_pre_jens}). This method was extended by Jeschke \textit{et al} \cite{Jesh03} to take into account the amplitude of ocean waves. The main drawback with this approach is encountered when trying to animate foam, since its motion does not follow the waves. Another problem arises from the empirical models that do not consider atmospheric conditions which are, in reality, the main cause of the apparition of foam. Oceanographic observations have shown for example that foam only appears when wind velocity exceeds $13$km/h \cite{Munk}. 

Premoze and Ashikhmin \cite{Pre01} compute the amount of foam based on wind velocity and the temperature difference between water and air. The user can control atmospheric conditions to induce more or less foam. However, results are limited by several constraints on the surface, and the animation is not visually convincing.
Darles \textit{et al} \cite{DCG07} extend this method with an animated texture, combined with other phenomena such as light attenuation due to spray particles in the air.

    \subsubsection{Particle systems}
      \begin{figure} 
      \begin{center}
      \includegraphics[width=7cm]{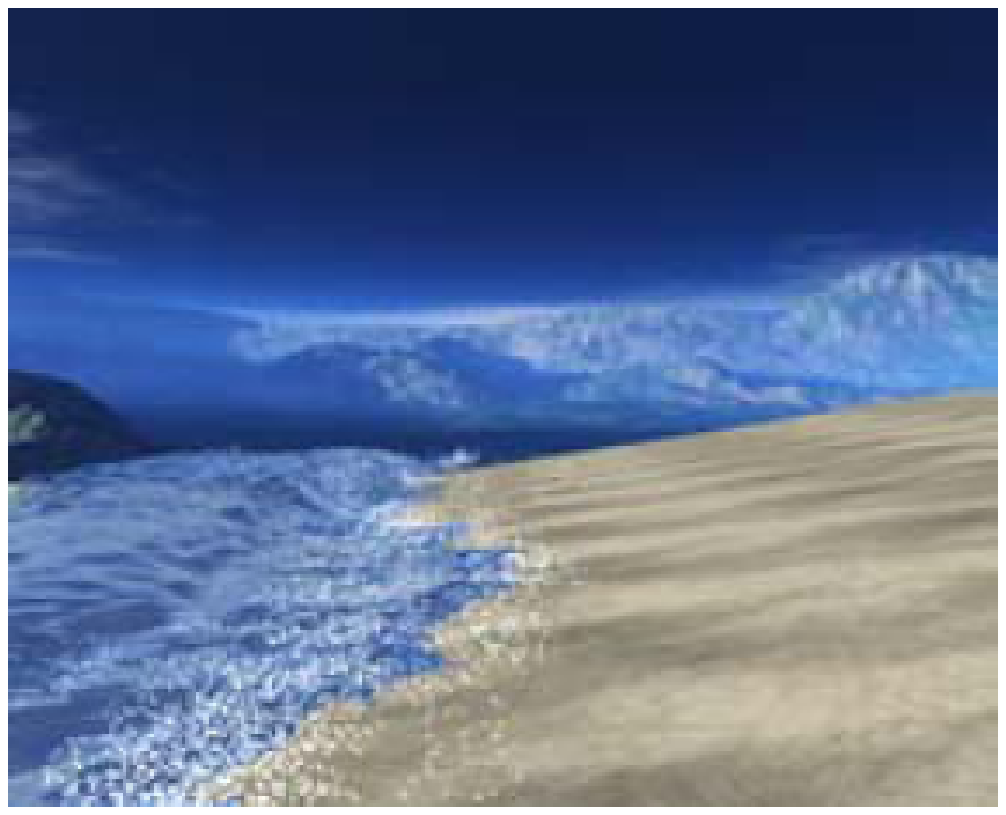}
      \end{center}
      \caption{Spray rendering from \cite{Wang06}}
      \label{fig:embruns marins_Wang06}
      \end{figure}
Peachey \cite{Peachey} first proposed to use particle systems in order to represent spray generated by breaking waves. Wang \textit{et al} \cite{Wang06} extend this approach by describing breaking waves using a Lagrangian approach, and generating spray subsystems according to the velocity of water particles \color{black}(see Figure \ref{fig:embruns marins_Wang06})\color{black}. Holmberg and Wunsche \cite{Holm04} generate foam particles by considering the amplitude of the waves, and render them as cloud points. The same idea is used by other authors \cite{Jens01, Jesh03, Chiu06} to generate spray particles according to the temporal variation of height at a given point. 
Although these methods provide realistic results, they also demand high computations since a large number of particles is needed; this proves difficult to handle when rendering large oceanic scenes.


    \subsection{Light-water interactions} \label{echanges_lumineux}
    \label{subsection:light_water}   

\color{black}
Light-water interactions play an important role in the way we perceive ocean's color, however they must be approximated especially for real-time implementations. We use their order of approximation to classify existing methods, from strong simplifications of optics laws to highly accurate computations generating rippling caustics formed when light shines through ocean waves.

\color{black}

    \subsubsection{First order approximation}
A first approach consists in rendering light-water interaction by applying geometrical optics laws, using Fresnel coefficients or \color{black}Schlick\color{black}'s approximation \cite{Schlick94}. The Beer-Lambert law reduces the intensity $I$ of a light ray by an exponential factor, depending on the \color{black}covered distance $d$ (\textit{i.e.} the distance traveled by the ray from the light source to the intersection point) \color{black} and an attenuation coefficient $a$:
    \begin{equation}
    I = I(0)e^{-ad}
    \end{equation}
    where $I(0)$ is the initial intensity of the ray.
In the case of reflexion, intensity can be reduced according to the optical depth due to particles in the air. For refraction, the reduction comes from the attenuation coefficient of water, as implemented by Tessendorf \cite{Tess01}.

Gonzato and Le Saec \cite{Gonz00} use Ivanoff's table defining the color of an ocean depending on its position on Earth \cite{JER74}. This color takes into account reflected and refracted light, and in presence of submarine objects, a fast light absorption method is applied by using an ocean water absorption spectrum.

In order to reduce computation costs and obtain realistic results in real-time, simpler methods and GPU implementations are often preferred.
An environment mapping method can be enough to approximate reflexion and refraction effects. In both cases, the viewer is assumed to be far from the considered point, thus incident rays are considered parallels and reflected rays only depend on the surface's normal vector. This technique is commonly used in computer graphics, especially for ocean surface rendering \cite{Schneider01, Jens01}. Baboud and Decoret \cite{BD06} represent refraction by an environment map of the bottom of the sea. An alternative method was proposed by different authors \cite{Bel03,Hu06} by mapping a sky texture onto the surface and considering this surface as perfectly specular.

\color{black}
Recently Bruneton et al. \cite{BNH10} proposed a GPU implementation of ocean lighting based on a hierarchical representation mixing geometry, normals and BRDF. 
The reflected light coming from the sun and the sky and the light refracted by the ocean are simulated by combining the BRDF model of Ross \textit{et al.} \cite{RDP05} and wave slope variance (\textit{i.e.} normals) depending on the required level of detail.
\color{black}

    \subsubsection{Multiple order approximation}
When a light ray enters the ocean surface, a fraction is diffused, absorbed and reflected on organic particles in suspension near the surface. In order to represent the complexity of this phenomenon and compute the amount of light at any point inside the water volume, a Radiative Transfer equation (RTE) must be solved.

     \begin{figure} 
    \begin{center}
    \includegraphics[width=7cm]{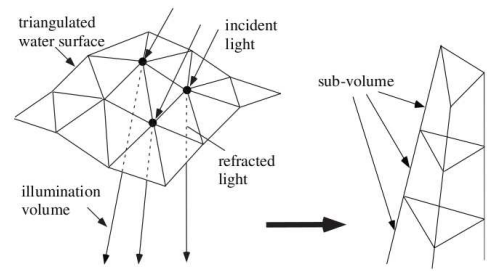}
    \end{center}
    \caption{Creation of illumination volumes from \cite{Iwa02}}
    \label{fig:vol_illum}   
\end{figure}
 
\color{black}
Arvo \cite{Arvo86} propose to emit rays from light sources and to accumulate them in illumination maps to represent caustics.
Iwasaki \textit{et al} \cite{Iwa01,Iwa03} use beam-tracing methods \cite{Heck84,Watt90}  to solve RTE by discretizing the water volume into \textit{illumination volumes} formed by the intersection of refracted rays and the bottom of the sea, sliced into sub-volumes (see Figure \ref{fig:vol_illum}). The amount of light bouncing back to the surface is computed by accumulating light in all traversed sub-volumes. By discretizing the water volume into horizontal, planar sheets, bouncing light can be computed by accumulating light at intersection points between rays and each sheet to obtain caustics. The photon mapping method proposed by Jensen \cite{Jensen01} can also simulate caustics or other global illumination effects.

Nishita \textit{et al} \cite{Nishita93} proposed a rendering algorithm to visualize oceans viewed from space by taking into account optical properties due to
atmospheric conditions. The color of the ocean is computed using a spectral model based on the RTE which takes several physicals parameters into account such as ocean depth, incident angles of the sun and viewing direction. 
\color{black}

      \begin{figure} 
      \begin{center}
      \includegraphics[width=7cm]{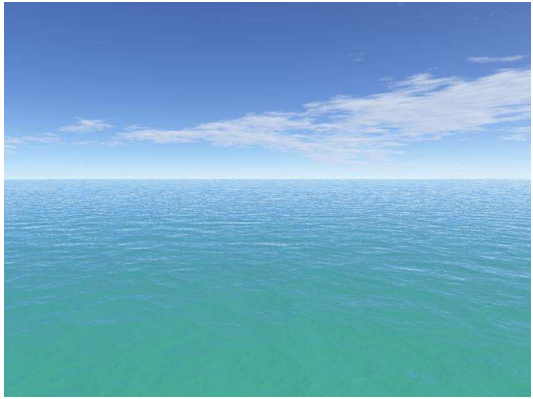}
      \end{center}
      \caption{Ocean surface rendering for a clear water from \cite{Pre01}}
      \label{fig:diffusion_premoze}
      \end{figure}

Sub-surface scattering was also addressed by several authors \cite{Pre01,Cerezo04} by taking into account several bio-optical parameters such as chlorophyll concentration and turbidity \cite{Jerlov}. In that case, coefficients for attenuation $a(\lambda)$ and diffusion $b(\lambda)$ are obtained from physical laws \cite{Morel91, Morel83} according to the phytoplankton concentration  $C_p$ inside water:
     \begin{eqnarray*} \label{coef_att}
      &a(\lambda)=&(a_{w}(\lambda)+0.06C_p^{0.65})(1+0.02e^{-0.014(\lambda-380)})\\
      &b(\lambda)=&\frac{550}{\lambda}0.30C_p^{0.32}
     \end{eqnarray*}
      where $a_{w}(\lambda)$ is the attenuation coefficient of pure water for a given wavelength $\lambda$. Thus turbidity is directly related to the concentration of phytoplankton. This approach is able to represent realistic ocean colors according to their biological properties (see Figure \ref{fig:diffusion_premoze}).

Gutierrez \textit{et al} \cite{Gutierez08} simulate a wider range of phenomena such as inelastic diffusion of light rays, fluorescence effects due to phytoplankton particles and Raman diffusion, observed when an incident ray's wavelength is modified because of water properties. 
This model is able to characterize the visual impact of different kinds of sediments on the resulting image, and thus represent different ocean waters depending on their bio-chemical composition.

Finally another type of methods consider subsurface scattering inside water as participating media. Interested readers should refer to \cite{PPS97} for an overview of these methods.

\color{black}\subsection{Discussion}

Numerous methods are available in the literature to take different optical phenomena into account and increase the realism of oceanic scenes. 
In the case of foam and sprays, empirical methods can effectively simulate these phenomena according to the state of disturbance of the surface. Combined with particle systems, these approaches allow to obtain realistic details, however they require a large number of particles which induces an important memory cost and computation time at rendering stage. Moreover, empirical methods are usually efficient for deep water scenes only; to our knowledge there is no work dealing with realistic foam created by breaking waves.\\
For light-water interactions, different methods were proposed to represent the complexity of these non-linear phenomena, yielding more and more realistic results.
Moreover, recent approaches even consider physical and biology effects to simulate a wide variety of virtual environments. All these works are summarized on Table \ref{fig:class3}.

\color{black}

\begin{table*}
\begin{center}
\begin{tabular}[t]{|l|c|c|c|c|c|}
  \hline
\multicolumn{2}{|c|}{} 	&\multicolumn{2}{c|}{Foam and Spray}	&	\multicolumn{2}{c|}{Light-water interactions}		\\
  \cline{3-6}
\multicolumn{2}{|c|}{} 	&Empirical	&Particles	&First order	&Multiple order	\\
  \hline
1986	&Arvo	\cite{Arvo86}	&	&	&	&x	\\
  \hline
1986	&Peachey	\cite{Peachey}	&	&x 	&	&	\\
  \hline
1993	&Nishita et al.	\cite{Nishita93}	&	&	&x 	&	\\
  \hline
2000	&Gonzato et al.	\cite{Gonz00}	&	&	&x 	&	\\
  \hline
2000	&Premoze et al.	\cite{Pre01}	&x 	&	& 	&x	\\
  \hline
2001	&Iwasaki et al.	\cite{Iwa01}	&	&	&	&x	\\
  \hline
2001	&Jensen	\cite{Jensen01}	&	&	&	&x	\\
  \hline
2001	&Jensen et al.	\cite{Jens01}	&x 	&x 	&x 	&	\\
  \hline
2001	&Schneider et al.	\cite{Schneider01}	&	&	&x 	&	\\
  \hline
2001	&Tessendorf	\cite{Tess01}	&	&	&x 	&	\\
  \hline
2003	&Belyaev	\cite{Bel03}	&	&	&x 	&	\\
  \hline
2003	&Iwasaki et al.	\cite{Iwa03}	&	&	&	&x	\\
  \hline
2003	&Jeschke et al.	\cite{Jesh03}	&x 	&x 	&	&	\\
  \hline
2004	&Cerezo et al.	\cite{Cerezo04}	&	&	&	&x	\\
  \hline
2004	&Holmberg et al.	\cite{Holm04}	&	&x 	&	&	\\
  \hline
2006	&Chiu et al.	\cite{Chiu06}	&	&x 	&	&	\\
  \hline
2006	&Baboud et al.	\cite{BD06}	&	&	&x 	&	\\
  \hline
2006	&Hu et al.	\cite{Hu06}	&	&	&x 	&	\\
  \hline
2006	&Wang et al.	\cite{Wang06}	&	&x 	&	&	\\
  \hline
2007	&Darles et al.	\cite{DCG07}	&x 	&	&x 	&	\\
  \hline
2008	&Gutierrez et al.	\cite{Gutierez08}	&	&	&	&x	\\
  \hline
2010	&Bruneton et al.	\cite{BNH10}	&	&	&x	&	\\

  \hline
\end{tabular}
\end{center}
\caption{Classification of the realistic rendering methods presented in section \ref{section:rendu}}
\label{fig:class3}
\end{table*}

\section{Conclusion}

In this survey we have presented available methods in computer graphics for modeling and rendering oceanic scenes. The great amount of different works is inherent to the extreme diversity of the phenomena involved. 

The first two sections of our survey focused on computer graphics models able to simulate the dynamical behavior of the ocean. We have seen that those methods are divided into two categories: on the one hand, \color{black}methods \color{black} usually dedicated to deep-water simulation, on the other hand \color{black}fluid-based approaches \color{black} trying to represent breaking waves near the shore. In the last section, we have seen different methods able to represent several phenomena involved in ocean rendering, namely foam, sprays and light-water interactions, that make for visual realism.

The next decade could see new methods emerging in an attempt to bridge the gap between deep-sea simulations and breaking waves representations; it would be necessary to put together different types of simulations on different scales. Scalable methods are also required for modeling and rendering stages in order to obtain real-time rates, which could include dynamic sampling of the simulation domain (as proposed by Yu \textit{et al} \cite{YNBH09} for rivers) and adaptive rendering models taking the visual impact of the considered phenomena into account.

\section*{Acknowledgements}

The authors thank C. Schlick and C. Rousselle for their careful proof reading. The authors also wish to
thank the reviewers for their careful reading and comments about earlier drafts of this paper.

\bibliographystyle{eg-alpha}
\bibliography{biblio}

\end{document}